\documentclass[preprint2]{aastex}

\usepackage{amsmath} 
\usepackage{amssymb} 
\usepackage{graphicx}
\usepackage{textcomp}
\usepackage{amsthm}
\usepackage{gensymb}
\usepackage{natbib}

\slugcomment{ }

\shorttitle{Improved Predictions of PBH Microlensing Rates}
\shortauthors{A. M. Cieplak & K. Griest}

\begin{document}


\title{Improved Theoretical Predictions of Microlensing Rates for the Detection of Primordial Black Hole Dark Matter}


\author{Agnieszka M. Cieplak and Kim Griest}
\affil{Department of Physics, University of California,
    San Diego, CA 92093}



\begin{abstract}
Primordial Black Holes (PBHs) remain a Dark Matter (DM) candidate of the Standard Model of Particle Physics. Previously, we proposed a new method of constraining the remaining PBH DM mass range using microlensing of stars monitored by NASA's Kepler mission. We improve this analysis using a more accurate treatment of the population of the Kepler source stars, their variability and limb-darkening. We extend the theoretically detectable PBH DM mass range down to $2\times10^{-10} M_\sun$, two orders of magnitude below current limits and one third order of magnitude below our previous estimate. We address how to extract the DM properties such as mass and spatial distribution if PBH microlensing events were detected. We correct an error in a well-known finite-source limb-darkening microlensing formula and also examine the effects of varying the light curve cadence on PBH DM detectability. We also introduce an approximation for estimating the predicted rate of detection per star as a function of the star's properties, thus allowing for selection of source stars in future missions, and extend our analysis to planned surveys, such as WFIRST.
\end{abstract}





\section{Introduction}

Dark Matter (DM) has been the topic of extensive research and remains one of the long standing mysteries in cosmology. Recent technological developments have aided the search for DM particle candidates \citep{Feng2010} with little success, so increased attention is now focused on closing the window of one of the few DM candidates left in the Standard Model of particle physics - primordial black holes (PBHs) \citep{Frampton2010, Carr2010}. As opposed to regular black holes, PBHs are much smaller and would only be able to form during the early universe, when perturbations could collapse to form stable PBHs whose mass would be on the order of the mass of the horizon at the time of their collapse. PBHs can form due to density fluctuations during different inflation scenarios, such as double inflation \citep{Frampton2010}, as well as due to phase transitions in the early universe causing a soft equation of state, bubble collisions, collapse of cosmic loops, or domain walls \citep{Khlopov2010}. First proposed by \citet{Zeldovich1966}  and \citet{Hawking1971}, PBHs would form during the radiation-dominated era, and therefore would be non-baryonic, satisfying the big bang nucelosynthesis limits on baryons, and would thus be classified as cold dark matter in agreement with the current paradigm. 

The discovery of Hawking evaporation \citep{Hawking1974, Hawking1975}, led to a theoretical lower limit on their mass scale of $5 \times10^{-19} M_\sun$, with any black holes smaller than this having evaporated by the current age of the Universe. There are no theoretical limits on the rest of the mass range, however higher masses have been progressively ruled out by various experiments \citep{Carr2010, Alcock1998}, leaving one major unconstrained window left, covering $5 \times 10^{-14} M_\sun$ to $2  \times10^{-8} M_\sun$. The lower mass end of this window is set due to femtolensing observations of gamma-ray bursts (see \citet{Barnacka2012} for a recent analysis of the Fermi Gamma-ray Burst Monitor data), while the higher mass end is constrained by the combined MACHO/EROS results due to microlensing \citep{Alcock1998}. In \citet[hereafter Paper I]{Griest2011}, we proposed to extend these microlensing constraints using the NASA Kepler satellite, which has the capability to close a significant part of the window. 

The Kepler satellite is a 1 m aperture telescope with a $115$ deg$^2$ field-of-view in an Earth trailing heliocentric orbit (see \citet{Koch2010, Borucki2010} for a description of the Kepler mission). It takes photometric measurements of around 150,000 stars every 30 minutes in the Cygnus-Lyra region. The telescope was launched in March 2009 and the mission has recently been extended to the year 2016. With planet-finding as its main science interest, it measures changes in stellar flux down to one part in a thousand or lower. This proves very beneficial in microlensing searches as well, where flux magnification is measured in the stellar light curves. 

Here we fill in details and expand on our previous analysis of this exciting possibility, analyzing the importance of limb-darkening on tightening our constraints, as well as calculating the probability for various lens parameters if any events are detected. We also improve our numerical estimate over the previous use of a 5000 star sample, by including all the third quarter Kepler stars being monitored ($\sim$156,000 stars), and project this as representative of the majority of the mission. We introduce new notation for the limb-darkened microlensing curves, which will be used to fit any future observed events. We correct an error in a well-known finite-source limb-darkening microlensing formula and derive a limb-darkened microlensing detection efficiency in our framework. Finally, we show that the PBH DM mass window can be extended further to lower masses using this improved analysis, and provide an approximation which can be used to predict microlensing rates in future surveys, such as the planned WFIRST space mission.

\section{Formulas}

\subsection{Point-Source Microlensing}

\citet{Paczynski1986} first proposed microlensing as a way to search for DM in the Milky Way. In doing so, he assumed a point-source, point-lens model, in which a lens, such as a PBH, would cause a magnification of this source when crossing in front of it, described by
\begin{equation}
A_{\mathrm{ps}} = \frac{u^{2} + 2}{u \sqrt{u^{2}+4}}
\label{eq:pointsource}
\end{equation}
where $u = b / r_{\mathrm{E}} $ and $b$ is the impact parameter of the lens, that is the transverse distance between the lens and the line-of-sight to the source. The Einstein ring radius $r_{\mathrm{E}}$ is given by
\begin{equation}
r_{\mathrm{E}}=\sqrt{\frac{4GMLx(1-x)}{c^{2}}}
\end{equation}
\noindent where $x$ is the ratio of the lens distance to the source distance, $L$ is the distance to the source star, and $M$ is the mass of the lens. As the PBH passes in front of the star, the amplitude becomes time-dependent, $A(t)=A[u(t)]$, and
\begin{equation}
u(t)=\left\{ u^{2}_{\mathrm{min}}+\left[\frac{2(t-t_{0})}{t_\mathrm{E}}\right]^{2}\right\}^{1/2}.
\end{equation}
Here $t_{0}$ is the time of the peak magnification, $u_{\mathrm{min}}=u(t_{0})$, and $t_\mathrm{E}=2r_{\mathrm{E}}/v_{\mathrm{t}}$ is the time for a lens to cross the Einstein ring with a velocity $v_\mathrm{t}$ transverse to the line of sight.

This is the standard point-source limit, in which the lens produces a 34 percent magnification when it is within one Einstein radius of the projected source star. This approximation is valid for a source that is much smaller than the Einstein radius and is not directly aligned with the lens. However, for the relatively nearby Kepler source stars and the relative low mass PBHs, the projected star radius needs to be taken into account.

\subsection{Finite-Source Microlensing}

When the projection of the radius of the star is comparable to the lens impact parameter, one needs to take into account finite-source effects on the detectability of events. The projected star radius is given by
\begin{equation}
U_*=\frac{R_* x}{r_\mathrm{E} (x)},
\end{equation}
\noindent where $R_*$ is the radius of the source star. For a constant surface brightness, equation \ref{eq:pointsource} now becomes \citep[eqns 9-11]{Witt1994}
\begin{equation}
A_{\mathrm{fs}}(U_*)=\frac{2}{\pi}+\frac{1+U^{2}_*}{U^{2}_*}\left(\frac{1}{2}+\frac{1}{\pi}\arcsin{\frac{U^{2}_*-1}{U^{2}_*+1}}\right)
\end{equation}
for $ u = U_*$, and 
\begin{align}
A_{\mathrm{fs}}(u,U_*) & =  \frac{2(u-U_*)^{2}}{\pi U^{2}_* (u+U_*)}\frac{1+U^{2}_* }{\sqrt{4+(u-U_*)^{2}}}\Pi \left(\frac{\pi}{2},n,k \right)  \nonumber 
\\ &+ \frac{u+U_* }{2 \pi U^{2}_* } \sqrt{4+(u-U_*)^{2}} E \left(\frac{\pi}{2},k \right) \nonumber
\\ &- \frac{u-U_* }{2 \pi U^{2}_* }\frac{8+(u^{2}-U^{2}_*) }{\sqrt{4+(u-U_*)^{2}}}F \left(\frac{\pi}{2},k \right) 
\end{align}
for $u \neq U_*$, where
\begin{equation}
n = \frac{4uU_*}{(u+U_*)^{2}}
\end{equation}
and
\begin{equation}
k = \sqrt{\frac{4n}{4+(u-U_*)^{2}}}.
\end{equation}
$F$, $E$, and $\Pi$ are elliptic integrals of the first, second, and third kind. As opposed to the point-source approximation, there is now a maximum amplitude for the magnification, which is equal to \citep[eqn 13]{Witt1994}
\begin{equation}
A_{\mathrm{max}_{\mathrm{fs}}} = \frac{\sqrt{4+U^{2}_*}}{U_*}.
\end{equation}

The duration of the event $t_{\mathrm{event}}$ will now be the time during which the event is detectable, starting when the lens crosses the threshold impact parameter $u_{\mathrm{thresh}}$ at which the microlensing light curve magnification reaches the minimum detectable threshold $A_{\mathrm{thresh}}=A(u_{\mathrm{thresh}},U_*)$. The duration of the event is then described by
\begin{equation}
t_{\mathrm{event}}=\left(u_{\mathrm{thresh}}^2 - u_{\mathrm{min}}^2\right)^{1/2} t_\mathrm{E}.
\end{equation}

As the ratio of the distance to the lens to the distance to the star, $x$, approaches $1$, the projected star radius in terms of the Einstein radius, $U_*$, approaches $\infty$, thereby suppressing the maximum magnification $A_{\mathrm{max}_{\mathrm{fs}}}$. Therefore there is some $x_{\mathrm{max}}$ beyond which $A_{\mathrm{max}_{\mathrm{fs}}}$ is lower than the detectable magnification, $A_{\mathrm{thresh}}$ and no events are detected. 
This  effect decreases the detection efficiency of the PBHs. However, the duration of the event increases, since the PBH does not have to be one Einstein radius away from the center, but from the edge of the projected star, for the beginning of a microlensing event. This effect increases the detection efficiency. As seen in Paper I, this dominates the detectability of events, increasing the number of expected microlensing events in the Kepler light curves. This finite-source model, however, assumes a constant brightness of the star, and does not take into account limb-darkening of the source star. In this paper we extend our analysis of the detectability of events to include this more physical model of the stars being lensed.

\subsection{Finite-Source Microlensing with Limb-Darkening}
\label{sec:Sec2.3}

For microlensing of nearby Kepler stars, where the Einstein radii of detectable PBHs is very small, limb-darkening is anticipated to be an important effect on the rate of detection. The effect is such that the star appears to be brighter towards the center, producing a  more concentrated source brightness, mimicking a model in between the point-source and the finite-source approximations. The limb-darkening profile we use to study this is the linear limb-darkening described by Witt and Mao, projected into the lens plane where \citep[eqn 15]{Witt1994}

\begin{equation}
I_{\mathrm{b}}(U_*^{'}) = 1- u_\lambda + u_\lambda \sqrt{1-(U_*^{' }/U_*)^2}.
\end{equation}

\noindent Here, $U_*^{'} $ is now the distance from the center of the projected star in terms of the Einsten radius and $u_\lambda$ is the linear limb-darkening coefficient. Witt and Mao calculated a limb-darkened profile numerically using a weighted surface brightness \citep[eqn 16]{Witt1994},

\begin{align}
A_{\mathrm{limb}}(u,U_*) = &\left(\int_0^{U_*} 2 \pi I_{\mathrm{b}}(U_*^{'}) dU_*^{'} \right)^{-1} \nonumber
\\& \times \int_0^{U_*}\frac{ \partial( A_{\mathrm{fs}}(u,U_*^{'})  \pi U_*^{' 2})}{\partial U_*^{'}}  I_\mathrm{b}(U_*^{'}) dU_*^{'}.
\label{eq:limb}
\end{align}

The integrand in the second integral of equation \ref{eq:limb} has a peak at  $u = U_*^{'}$, which causes some problems with convergence when integrating numerically. Witt and Mao provided another form of equation \ref{eq:limb}, \citep[first half of eqn 16]{Witt1994} but recommended against its use due to, they said, the presence of a singularity. We find in fact, that there is an error in their equation, due to the treatment of the integral boundaries. Integrating equation \ref{eq:limb} by parts we find the correct magnification for a linear limb-darkened profile to be

\begin{align}
& A_{\mathrm{limb}}(u,U_*) = (1-u_{\lambda}/3)^{-1} \left[ \vphantom{ \int_0^{U_*} }A_{\mathrm{fs}}(u,U_*) \pi U_*^{2} (1-u_{\lambda}) \right.\nonumber
 \\& \qquad \left. +(\pi u_{\lambda} / U_*) \int_0^{U_*} A_{\mathrm{fs}}(u,\sqrt{U_*^{2}-z^{2}}) (U_*^{2}-z^{2}) dz \right],
 \label{eq:newlimb}
\end{align}
where
 \begin{equation}
A_{\mathrm{fs}}(y) \pi y^{2} =2 y+(1+y^{2}) \left(\frac{\pi}{2}+\arcsin{\frac{U^{2}_*-1}{U^{2}_*+1}} \right)
\end{equation}
for $ u = y$, and 
\begin{align}
A_{\mathrm{fs}}(u,y) \pi y^{2} & =  \frac{2(u-y)^{2}}{(u+y)}\frac{1+y^{2} }{\sqrt{4+(u-y)^{2}}}\Pi \left(\frac{\pi}{2},n,k \right)  \nonumber 
\\ &+ \frac{u+y}{2} \sqrt{4+(u-y)^{2}} E \left(\frac{\pi}{2},k \right) 
\\ &- \frac{u-y}{2}\frac{8+(u^{2}-y^{2}) }{\sqrt{4+(u-y)^{2}}}F \left(\frac{\pi}{2},k \right) \nonumber
\end{align}
for $u \neq y$.
Here $y = U_*$ or $y = \sqrt{U_*^{2}-z^{2}}$, where appropriate in equation \ref{eq:newlimb}. Use of this equation removes the problem of numerical convergence in equation \ref{eq:limb}.

As stated above, the limb-darkening of the source star produces a more concentrated source brightness, thereby changing the shape of the microlensing light curve. There will therefore be a higher maximum magnification than that produced by a pure finite-source light curve. \citet{Agol2002} calculated this maximum amplification for a quadratic limb-darkening model. Here we present the result for the linear limb-darkening profile using equation \ref{eq:newlimb} when $u=0$:

\begin{align}
& A_{\mathrm{max}}= \frac{A_{\mathrm{max}_{\mathrm{fs}}}}{1-u_\lambda/3} \nonumber
\\ & \times \left \{\vphantom{\frac{A_{max_{fs}}}{1-u_\lambda/3}}1-u_{\lambda}+(2 u_{\lambda}/ 3 U_*^{2}) \left[\vphantom{ \frac{A_{\mathrm{max}_{\mathrm{fs}}}}{1-u_\lambda/3} }(2+ U_*^{2})E \left[U_*^{2}/(4+U_*^{2}) \right]  \right . \right . \nonumber
\\ & \qquad \qquad \qquad \qquad \qquad  \left . \left . \vphantom{\frac{A_{\mathrm{max}_{\mathrm{fs}}}}{1-u_\lambda/3}} -2K \left[U_*^{2}/(4+U_*^{2}) \right] \right] \right \}  ,
\label{eq:amax}
\end{align}

\noindent where $K$ and $E$ are the complete elliptic integrals of the first and second kind.

Therefore, there is a new, higher $x_{\mathrm{max}}$ below which the magnification is detected, allowing more PBHs to be observed. At the same time, since the brightness of the star is more concentrated, the impact parameter for which an event is detected will be closer to the projected center of the star, decreasing the duration of the event. Since this is an effect in between the point-source and finite-source model, there should be an overall decrease in the number of expected events in the Kepler data.

\section{Effect of Limb Darkening on the Numerical Estimate of Expected Number of Events}
\label{sec:SecLimb}

\begin{figure}[htb!]
\begin{center}
\includegraphics[scale=0.5, trim=0.5in 0 0 0]{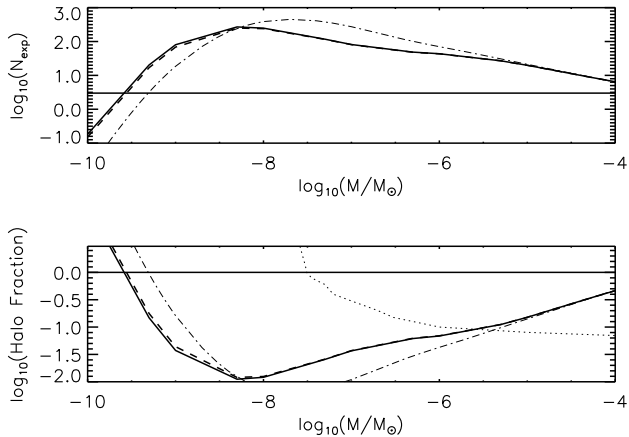}
\caption{\emph{Top panel:} Expected number of events, $N_{\mathrm{exp}}$, scaled to 780,000 star-years of Kepler observing time  for a finite-source microlensing model with no limb-darkening (solid line) and with limb-darkening (dashed line). We assume 4 sequential measurements with 3-sigma above average measurements of the flux. The dash-dotted line represents $N_{\mathrm{exp}}$ from our previous estimate of Paper I. The horizontal line shows the 95\% confidence level limit if no events are detected. \emph{Bottom panel:} The potential 95\% confidence level exclusion of PBH dark matter. The area above the solid line for the finite-source model (dashed line when limb-darkening is included) would be ruled out if no events are detected in the 7.5 year Kepler observation lifetime. The dash-dotted line represents the previous estimate of Paper I, while the dotted line represents the current limits from the combined MACHO/EROS LMC microlensing surveys \citep{Alcock1998}. The horizontal line depicts a DM halo consisting entirely of PBHs. }
\label{fig:Fig1}
\end{center} 
\end{figure}

In Paper I we calculated the expected number of PBH microlensing events in the Kepler data using a subset of 5000 stars of the third quarter light curves. Here we make a more accurate estimate of this number. We first extend the calculation to include the full set of the third quarter light curves being monitored, reflecting a more accurate sample of the stars. In addition, we scale the observation time to 7.5 years to include the extended Kepler mission, as well as assume a more accurate number of variable stars in the Kepler mission, with $25\%$ of the observed dwarf stars and $95\%$ of the observed giant stars assumed to be variable \citep{Ciardi2011}. This results in a total of 780,000 star-years being observed, twice the amount assumed in Paper I, however now the distribution is assumed as consisting of mainly dwarf stars with many of the giant stars being unusable. The previous estimate of the number of expected PBH microlensing events cannot therefore be naively multiplied by two. As in Paper I we require 4 sequential measurements 3-sigma above average, equivalent to a microlensing event of a minimum 2 hours duration.  Calculating the stellar distances and the magnification thresholds $A_{\mathrm{thresh}}$ (below which an event would not be detected) for each star as described in Paper I, we recalculate the number of expected events using equation 1 of Paper I, for a finite-source model of microlensing, with this new distribution of stars. This is plotted in Figure \ref{fig:Fig1} as the solid line.  The shape and peak of the number of events for each PBH mass has changed. Discarding $95\%$ of the giant stars as variable results in the curve being dominated by the dwarf stars' sensitivity. While the giant stars' peak PBH DM mass sensitivity is at around $10^{-8} M_{\sun}$, which dominated the earlier curve in Paper I, the dwarf stars' sensitivity is better at the lower PBH masses, peaking at around $5 \times 10^{-9} M_{\sun}$, since a higher fraction of the stellar intensity is magnified. The larger amount of star-years, dominated by monitoring of dwarf stars, increases the sensitivity to lower mass PBHs. This more accurate estimate shows a potential closing of the PBH DM window down to $2 \times 10^{-10} M_{\sun}$, compared to the $5 \times 10^{-10}$ previously estimated.

In addition to these calculations, we now consider the effect of limb-darkening on the predicted number of detectable events. As seen in the previous section, limb-darkening increases $A_{\mathrm{max}}$, but decreases the duration of the event, since the total stellar flux remains the same. $A_{\mathrm{max}}$ determines $x_{\mathrm{max}}$, the distance to which a PBH would be detectable, with $1$ being the maximum value. A higher $A_{\mathrm{max}}$ would naively increase this value, increasing the range of masses that PBHs would be detectable, however, calculating $x_{\mathrm{max}}$ for a typical Kepler star (with radii between $0.9 R_{\sun}$ and $1.5 R_{\sun}$), without limb-darkening, one can see that $x_{\mathrm{max}}$ is already approaching the maximum value of 1. Therefore, increasing this further when limb-darkening is added, does not produce much of an effect on the total number of events that could be detected. The only other effect then is to reduce the possible duration of an event, therefore decreasing the total expected number. In Paper I, we showed that the naive point source optical depth (the total number of PBHs inside a microlensing tube as defined in \citet{Griest1991}) is proportional to $u_{\mathrm{thresh}}^2$ (the detectable impact parameter value) and thus replacing this with the projected star radius, changes the optical depth by a factor of $U_*^2$. Extending this naive calculation to limb-darkening, we can see that any effect on the projected star radius will directly impact the optical depth in quadrature. Therefore, limb-darkening, which effectively reduces the radius of the star,  could drastically limit how far we can extend the current PBH mass range, since lower PBH masses would not be as detectable. 

Here we explore this effect by repeating the above calculation for the expected number of PBH microlensing events, however now including the limb-darkening effect for each star. In order to do this, we used the linear limb-darkening model as described in Section \ref{sec:Sec2.3}, calculating the linear limb-darkening coefficients using the \citep{Sing2010} model grid to find Kepler limb-darkening coefficients as a function of the effective temperature, surface gravity, and metallicity of each of the 150,000 Kepler source stars. This enables us to calculate a new detectable magnification threshold $A_{\mathrm{thresh}}$ using the magnification formula in equation \ref{eq:newlimb} for linear limb-darkening. The number of expected events including this linear limb-darkening effect is plotted in Figure \ref{fig:Fig1} as the dashed line.

\begin{figure}[htb!]
\begin{center}
\includegraphics[scale=.5,trim=0.5in 0 0 0]{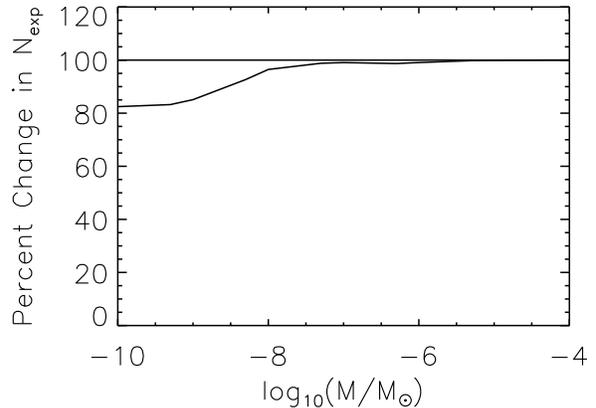}
\caption{Percent change in the number of expected microlensing events due to limb-darkening for 780,000 star-years of Kepler observing time, defined as $N_{\mathrm{exp}}$ with limb darkening  divided by $N_{\mathrm{exp}}$ without limb-darkening included. The horizontal line depicts 100\%, representing no change, for comparison.}
\label{fig:Fig2}
\end{center} 
\end{figure}

Surprisingly, Figure \ref{fig:Fig1} does not show a significant effect of limb-darkening on the number of expected events. This can be understood as being due to the extreme precision of Kepler. Since Kepler light curves allow a magnification threshold of $A_{\mathrm{thresh}}=1.001$ or lower to be set, the median Kepler star (with a small stellar radius), will allow for a detection as soon as a PBH is within one Einstein radius of the projected star radius, where the limb-darkening does not play much of an effect yet on the light curve. On stars with bigger stellar radii, as well as models with smaller PBH masses, this limb-darkening will have some effect, since the magnification reaches $A_{\mathrm{thresh}}$ only when the PBH's impact parameter is inside the projected star radius, where the limb-darkening plays a role. However, even then, the duration is shortened by just a small amount.
 In Figure \ref{fig:Fig2} we plot the percent change between the finite-source model with and without limb-darkening and we in fact see that there is a decrease of up to $17\%$ for the lower lens masses. Moving towards higher PBH masses, almost all the Kepler stars which have radii on the order of a solar radius, will be able to have a detectable event within one Einstein radius of their projected radius, therefore, making the limb-darkening effects negligible. We found that if we did not discard $95\%$ of the giant stars as variable, we would see an increase in the overall number of expected events  due to limb-darkening at higher mass ranges. $A_{\mathrm{max}}$ is significantly lower for stars with large radii, and therefore the detectable value of $x$ will be close to $0$. We found that introducing limb-darkening in giant stars increases this $A_{\mathrm{max}}$, increasing the possible detectable range of $x$, drastically increasing the number of detectable events. Also, the higher the mass of the PBH, the lower the projected star radius, and the more pronounced this effect will be, increasing the $x$ ratio by a higher amount when introducing limb-darkening. Therefore the biggest effect of limb-darkening on the number of events seems to be in these large-radii stars and on the lower lens masses. The effect of limb-darkening will also be crucial in the fitting of potential PBH events as well as in calculating the experimental detection efficiency.

\section{Detection Efficiency}

In order to turn either detection or non-detection of microlensing into a statement about Galactic dark matter, one must calculate the efficiency at which the experiment finds PBH microlensing. This is done by performing a Monte Carlo simulation with randomly generated microlensing events and using the same selection criteria used to select microlensing light curves. One thus calculates the detection efficiency $\epsilon$, which is just the fraction of simulated events recovered. Using $\epsilon$ one can calculate the expected number of detectable events as 

\begin{align}
N_{\mathrm{exp}} = & \int_0^{x_{\mathrm{max}}} \int_0^{u_{\mathrm{thresh}}(x)} \int_0^{\infty}  \epsilon(x,u_{\mathrm{min}},v_\mathrm{t}) \nonumber
\\ & \times  \frac{d\Gamma}{dx du_{\mathrm{min}} dv_\mathrm{t}}dv_\mathrm{t} du_{\mathrm{min}} dx,
\end{align}
where
\begin{equation}
 \frac{d\Gamma}{dx du_{\mathrm{min}} dv_\mathrm{t}} = 4 r_\mathrm{E}(x) L \frac{\rho}{M} \frac{v_\mathrm{t}^2}{v_\mathrm{c}^2} e^{-v_\mathrm{t}^2/v_\mathrm{c}^2}  ,
 \end{equation}
 where $v_\mathrm{c} \approx 220$ km/s is the halo circular velocity, $\rho \approx 0.3$ GeV cm$^{-3} $ is the local dark matter density, and $v_\mathrm{t}$ is the transverse lens velocity. Here we are making the approximation that the DM density is constant between the Earth and the source stars, valid for the relatively nearby Kepler stars, and using the fact that the Kepler field is nearly in the direction of the Sun's motion around the galaxy (see Paper I).
 
 Without limb-darkening, $\epsilon$ is a function of only $x$ and $v_\mathrm{t}$, and the integral over $u_{\mathrm{min}}$ can be performed since that distribution is well-known due to a uniform stellar intensity across the projected star radius. This is not true when limb-darkening is included, since a lower $u_{\mathrm{min}}$ produces a higher amplitude in a limb-darkened light curve as well as decreases the duration of an event. The effect of limb-darkening on the efficiency is calculated by adding the parameter $u_{\mathrm{min}}$ to the Monte Carlo simulation. Thus the two competing effects can both influence this probability of detection.

\section{Detectable Parameters}

Also of interest is what we can say about a potential PBH if we do detect microlensing events. Assuming the measured light curve parameters are $U_*$ and $t_{\mathrm{event}}$, one can then calculate the mass probability and the distance probability of the lens in terms of these measured parameters. Performing a change of variables in terms of the observable parameters, the mass and distance likelihood functions can be derived. We find the mass likelihood function to be

\begin{equation}
\frac{d \Gamma}{dt_{\mathrm{event}} dU_*} = \frac{\rho v_\mathrm{c}^2 c^2}{2 G} \frac{R_*^2 U_*}{M^2} \left(\frac{c^2 R_*^2}{4 G M L}+U_*^2\right)^{-2} \beta^2 g(\beta),
\label{eq:masslikelihood}
\end{equation}
where
\begin{equation}
\beta=\frac{4 u_{\mathrm{thresh}}(U_*,u_\lambda)^2 R_*^2 U_*^2}{t_{\mathrm{event}}^2 v_c^2} \left(\frac{c^2 R_*^2}{4 G M L}+U_*^2\right)^{-2}
\end{equation}
and
\begin{equation}
g(\beta) = \int_0^1 dy y^{3/2} (1-y)^{-1/2} e^{-\beta y}.
\label{eq:masslikelihoodbeta}
\end{equation}

\begin{figure}[htb!]
\begin{center}
\includegraphics[scale=.5, trim= 0.5in 0 0 0]{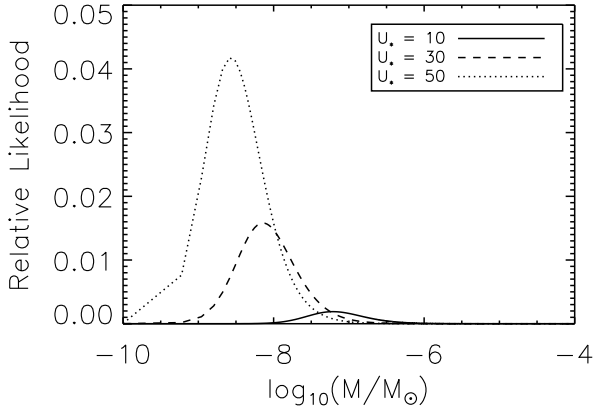}
\caption{Lens mass relative likelihood for  $t_{\mathrm{event}}=2$ hours, $R_*=1 R_{\sun}$, $L=0.73$ kpc, $u_\lambda = 0.61$, and $A_{\mathrm{limb}}(u_{\mathrm{thresh}},U_*) = 1.0007$ representing a median star in the Kepler field. The range of values for $U_{*}=10, 30$,  and $50$ represents the range that could be measured in such a star. The curves here are normalized to have unit area under each curve, so that each curve represents the probability of each mass given the measured lightcurve parameters.}
\label{fig:Fig3}
\end{center} 
\end{figure}

In Figure \ref{fig:Fig3} we plot this likelihood for $t_{\mathrm{event}}=2$ hours, $R_*=1 R_{\sun}$, $L=0.73$ kpc, $u_\lambda = 0.61$, and $A_{\mathrm{thresh}}=A_{\mathrm{limb}}(u_{\mathrm{thresh}},U_*) = 1.0007$, values typical of a median star in the Kepler field being monitored. The different curves represent the range of different values of $U_*$ that could be measured for such a star undergoing a microlensing event. The distributions are normalized to have unit area under each curve so that each curve can be thought of as a probability density, that is for a measured duration of 2 hours and the given $U_*$, the curves give the relative likelihood of the event being caused by a PBH of mass shown on the abscissa. The rise of each curve at low mass is dominated by the $\beta^2$ term, which increases with increasing $U_*$. The event duration is proportional to this projected star radius $U_*$, the PBH Einstein radius, $r_\mathrm{E}$, and inversely proportional to the PBH transverse velocity, $v_\mathrm{t}$. For a lower mass PBH only small transverse velocities will give rise to events longer than 2 hours. If $U_*$ is decreased, then for the lower mass PBHs the transverse velocities have to be smaller, with the number of possible events for these mass ranges approaching zero. This is the reason that the distributions tend to zero for the smaller values of $U_*$ for the smaller PBH masses in Figure \ref{fig:Fig3}. On the other hand, the decrease in each curve at larger PBH masses is caused by the lower number density of PBHs as their mass increases. The bigger the PBH lenses, the lower the number density needed to explain the local dark matter density. This in turn corresponds to less potential microlensing events. If microlensing events were to be observed in the Kepler data, we would then be able to, using distributions such as these, estimate the mass of the PBH making up the DM. The product of these likelihood functions would give us an estimate of the PBH DM mass range. We could also use these distributions to exclude some range of masses that the microlensing lenses could represent by measuring their $t_{\mathrm{event}}$ and $U_*$ parameters.

In a similar way, we can calculate the lens distance likelihood function, with a change of variables in Equations \ref{eq:masslikelihood} - \ref{eq:masslikelihoodbeta},

\begin{equation}
\frac{d \Gamma}{dt_{\mathrm{event}} dU_*} = 8 G \rho \frac{v_\mathrm{c}^2}{c^2} \frac{L^2}{R_*^2} U_* (1-x)^2 \beta^2 g(\beta),
\end{equation}

\noindent where

\begin{equation}
\beta = 4 \frac{R_*^2}{t_{\mathrm{event}}^2 v_\mathrm{c}^2} \frac{u_{\mathrm{thresh}}(U_*,u_\lambda)^2}{U_*^2} x^2.
\end{equation}

\begin{figure}[htb!]
\begin{center}
\includegraphics[scale=.5, trim=0.5in 0 0 0]{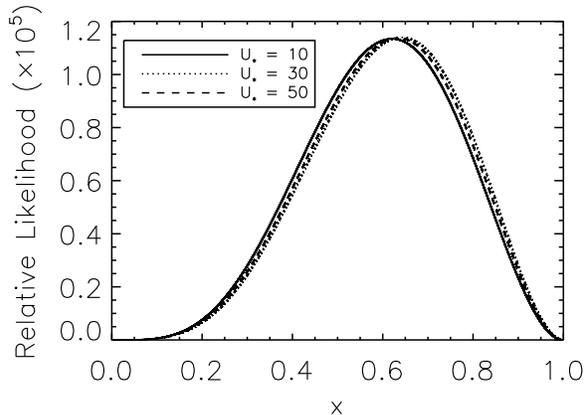}
\caption{Lens distance relative likelihood for  $t_{\mathrm{event}}=2$ hours, $R_*=1 R_{\sun}$, $u_\lambda = 0.61$, and $A_{\mathrm{limb}}(u_{\mathrm{thresh}},U_*) = 1.0007$ representing a median star in the Kepler field. The range of values for $U_{*}=10, 30$, and $50$ represents the range that could be measured in such a star. The curves here are normalized to have unit area under each curve.}
\label{fig:Fig4}
\end{center} 
\end{figure}

These distribution functions are plotted normalized to unit area under each curve in Figure \ref{fig:Fig4}. We see that the distance to the lens is not very dependent on the value of $U_*$ measured. The distance probability distribution is dominated by the transverse velocities that are detectable at each distance to the lens (closer lenses have to be traveling slower in order to be detected for a given measured $t_{\mathrm{event}}$). If a microlensing event was to be detected, then the probability distribution for the distance to the lens can be plotted for the particular stellar radius being monitored, just like in Figure \ref{fig:Fig4}. A bigger $R_*$ will yield a distribution centered at lower $x$, while a smaller $R_*$ will shift the distribution to higher $x$. We can then narrow down the most likely position of the lens based on the likelihood of detection at each distance for the particular star at which the event occurred. Overall, though, Figure \ref{fig:Fig4} shows that it will not be easy to determine the distances to the lens.

\section{Limitations}

\begin{figure}[htb!]
\begin{center}
\includegraphics[scale=.5,trim=0.5in 0 0 0]{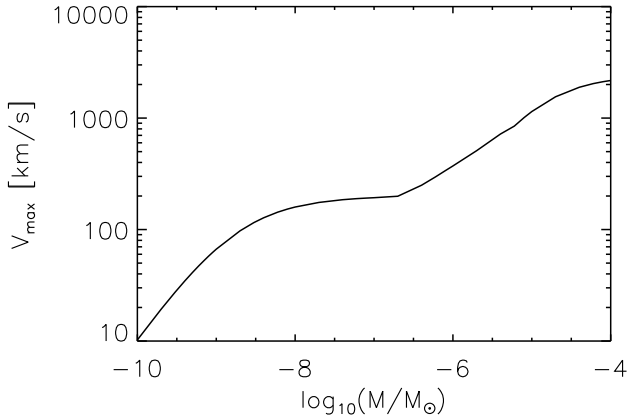}
\caption{Maximum detectable transverse velocity, $v_{\mathrm{max}}$, of PBHs, for a median type Kepler star with $R_*=1 R_{\sun}$,  $L=0.73$ kpc, $u_\lambda = 0.61$, and $A_{\mathrm{limb}}(u_{\mathrm{thresh}},U_*) = 1.0007$, for a minimum $t_{\mathrm{event}}=2$ hours.}
\label{fig:Fig5}
\end{center} 
\end{figure}

We turn now to considerations of what could be improved in upcoming missions, and what theoretical limitations the Kepler satellite has on detecting PBHs. As mentioned above, for the lower mass range of PBHs, the higher-velocity objects would not be detectable, because they result in magnifications that last too short a time. In Figure \ref{fig:Fig5} we plot the maximum detectable velocity for a given PBH mass for a Kepler star with median parameter values of distance, radius, and $A_{\mathrm{thresh}}$, calculated from the third quarter stars monitored. Throughout this section, by velocity we mean the velocity of the lens relative to the Earth-source line-of-sight in the direction perpendicular to the line-of-sight. For a median type Kepler star, the $v_\mathrm{c}$ value of $220$ km/s is on average detectable for masses above $2 \times10^{-7}$ solar masses. The lower-mass range events would be dominated by the projected radius of the star and therefore their maximum velocity curve approaches a constant value, but at the upper-mass range, the Einstein radii of the PBHs become important in detecting the event, and therefore more of the velocity distribution is detectable.

\begin{figure}[htb!]
\begin{center}
\includegraphics[scale=.5,trim=0.5in 0 0 0]{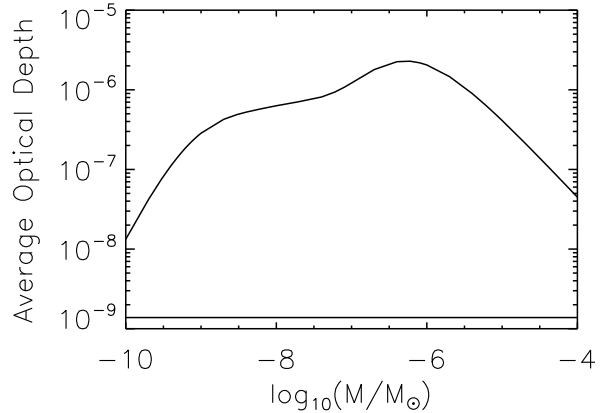}
\caption{Average optical depth for the third quarter Kepler source stars being monitored. The horizontal line depicts the average optical depth if a point-source microlensing model is used for comparison.}
\label{fig:Fig6}
\end{center} 
\end{figure}

In order to understand the limits on this detectable velocity, we calculated the optical depth, which determines how many measurements are needed for any given PBH mass, and is defined as the number of PBHs inside a detectable microlensing tube (as defined in Griest 1991, with the addition of an $x_{\mathrm{max}}$ cutoff for the finite-source model). We therefore arrive at the equation $\tau = \pi (\rho/M) L\int_0^{x_{\mathrm{max}}} u^2_{\mathrm{thresh}}(x) r_\mathrm{E}^2(x) dx$. In Figure \ref{fig:Fig6} we plot the optical depth, averaged over all the third quarter Kepler source stars. The overall amplitude of the curve is set by the average distance to the stars being monitored, whereas the shape of the curve at lower-mass PBHs is mostly governed by $x_{\mathrm{max}}$, the detectable distance to the lens. This is set by the photometric accuracy of Kepler and cannot be changed.

\begin{figure}[htb!]
\begin{center}
\includegraphics[scale=.5, trim=0.5in 0 0 0]{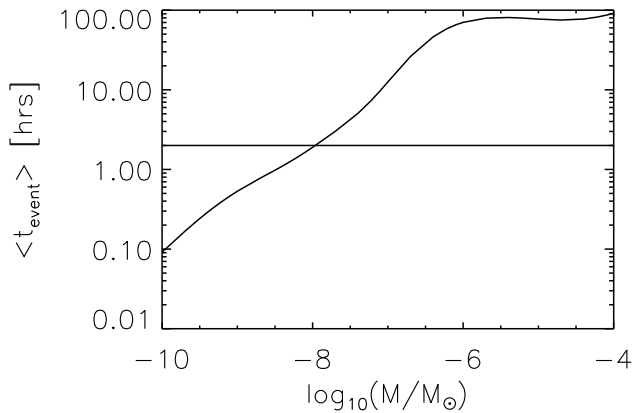}
\caption{Average event duration, $<t_{\mathrm{event}}>$, for the third quarter Kepler source stars being monitored. The horizontal line depicts $t_{\mathrm{event}}=2$ hours, which represents the minimum event duration required for a microlensing detection in the Kepler data.}
\label{fig:Fig7}
\end{center} 
\end{figure}

Another limitation is related to our detection of events that last 2 hours or more, which is set by the Kepler satellite cadence. In Figure \ref{fig:Fig7} we plot the average event duration, defined as $t_{\mathrm{event}}=\tau/\Gamma$, averaged over all the third quarter Kepler stars being monitored. We can see that the average event duration is about 2 hours at around $10^{-8}$ solar masses. With the 2 hour selection criteria, we are therefore not able to detect most events for the smaller mass PBHs. Figure \ref{fig:Fig7} shows that if events as short as 0.1 hours could be detected, one might be able to detect PBHs of masses down to $10^{-10} M_\sun$ or below. This could be improved upon by decreasing the Kepler satellite cadence when monitoring the stars. 

Our theoretical detectable limits are due to both the threshold detection limit of Kepler and the cadence. The first cannot be changed, however we address improvements to the second limitation in the next section.

\section{Future Planned Missions}

\begin{figure}[htb!]
\begin{center}
\includegraphics[scale=.5,trim=1in 0 0 0]{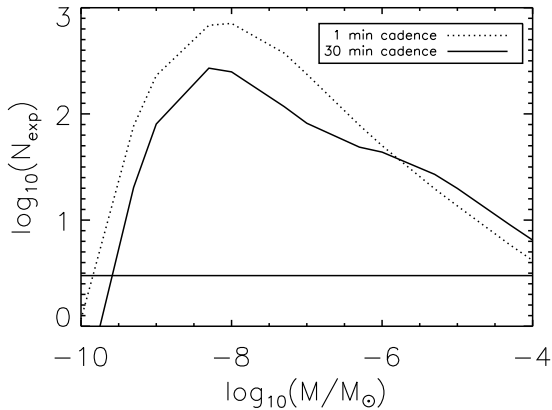}
\caption{Total number of expected events for 780,000 star-years for 1 min and 30 min Kepler cadence monitoring times. The horizontal line shows the 95\% confidence level limit if no events are detected.  }
\label{fig:Fig8}
\end{center} 
\end{figure}

Kepler's extended mission of four additional years helps increase its sensitivity to lower-mass DM PBHs, as seen in Section \ref{sec:SecLimb}. Here we address whether any possible additional measures could be undertaken during this extended mission to further increase this sensitivity. As seen in the previous section, in the analysis of the average microlensing event duration, cadence plays a huge role in these measurements. The Kepler camera actually takes one image every minute. Due to communication bandwidth, for most stars, the Kepler team adds up the one minute exposures into a 30 minute exposure before transmitting the data to Earth. These are called ``long cadence" light curves. However, for a selectable subset of stars, the entire one minute cadence (``short cadence") light curves are transmitted. Thus we wish to investigate the value for microlensing of the Kepler team returning short cadence data on a subset of stars. If we naively decrease the monitoring cadence to 1 min for all the Kepler source stars, for the full 780,000 star-years, the lower-mass DM PBH sensitivity increases by an order of magnitude, down to $2\times 10^{-11} M_\sun$. Thus, while not possible due to bandwidth limitations, naively returning short cadence light curves for all the stars, would allow the exploration of one additional factor of 10 in the allowed PBH mass range. This could be an exciting possibility. However, when the cadence is decreased, the Poisson average error in each flux measurement increases. Thus with the set light gathering power of the Kepler telescope, a shorter cadence is offset by a larger $A_{\mathrm{thresh}}$. In order to investigate this trade-off we redid our analysis assuming a 1 min cadence, but reducing the signal/noise for each flux measurement appropriately. Figure \ref{fig:Fig8} shows the results of this analysis for 780,000 star-years with the same assumed stellar variability as in Section \ref{sec:SecLimb}. Although requiring short cadence for all the Kepler stars for the full mission is not achievable, this demonstrates that reducing the cadence time on the current source stars with the current light curve precision, would give only a modest increase in the sensitivity to lower-mass PBHs. However, if Kepler's aperture was large enough to maintain the current signal/noise at a one minute cadence, great improvement would be possible. In this we note that certain stars are far more valuable than others for detecting microlensing.

\begin{figure}[htb!]
\begin{center}
\includegraphics[scale=.5, trim=0.5in 0 0 0]{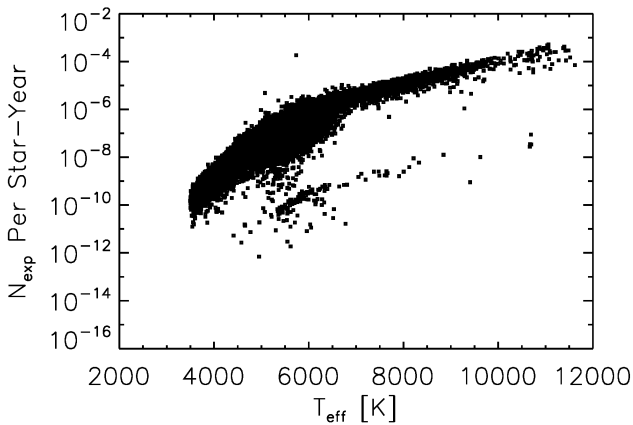}
\caption{Expected number of events per star-year for a PBH mass of $10^{-10} M_\sun$ with a 1 min cadence, for each non-variable dwarf star being monitored in the third quarter of the Kepler data, plotted with respect to $T_{\mathrm{eff}}$. }
\label{fig:Fig9}
\end{center} 
\end{figure}

We therefore address the characteristics of source stars and cadence times that would optimize this sensitivity in future missions similar to Kepler. Since 95\% of the giant stars are seen to be variable, we focus on the dwarf stars, which would yield a higher number of less variable light curves for the same number of stars monitored. To investigate an optimal selection of stars for monitoring we calculated the expected number of microlensing events for each non-variable dwarf star in the third quarter Kepler data. We plotted this per-star-$N_{\mathrm{exp}}$ versus other stellar parameters to see which correlated well with higher $N_{\mathrm{exp}}$. In Figure \ref{fig:Fig9}, we show the best such correlation, $N_{\mathrm{exp}}$ vs. $T_{\mathrm{eff}}$, where stars with higher $T_{\mathrm{eff}}$ are much more likely to return a detectable microlensing event. This figure shows that source stars with $T_{\mathrm{eff}}>8000K$ are roughly 100,000 times more valuable for monitoring than stars with $T_{\mathrm{eff}}<4000K$. Thus, short cadence monitoring of a handful of carefully selected stars should be as valuable as large numbers of typical stars. The bandwidth problem, therefore, might be solved by long cadence monitoring of most stars but short cadence monitoring of a small sample. 

Why does $T_{\mathrm{eff}}$ correlate so well with the expected number of lensing events? We see that for a point-source microlensing model detectability correlates well with luminosity; the more luminous stars give higher rates of predicted detections when the projected star radii and the lens Einstein radii are comparable. However, for these small PBH masses, the Einstein radii are much smaller than the projected star radii and the luminosity per unit area is more important, giving the upward trend with effective temperature.

In order to understand better the factors that influence the number of predicted events for a given star, we can make the following approximations. For light curves with 1\% or better precision, as seen in Kepler, we can approximate $u_{\mathrm{thresh}} \approx U_*$. Also, for  $R_*/R_\sun < 0.57 \left(t_{\mathrm{event}}/1hr \right) \left(v_\mathrm{c}/220 km/s \right)/x_{\mathrm{max}}$, we can approximate $\beta<1$ and $\beta^2 g(\beta) \approx (3/8)\pi \beta^2$. Using these approximations in equation 1 of Paper I, with $x_{\mathrm{max}}<1$ for $M = 10^{-10} M_\sun$, we arrive at the predicted rate of detection for a given star

\begin{equation}
\Gamma \approx 409.6 \pi \frac{G^5 M^4 \rho}{c^{10}v_\mathrm{c}^2}  \frac{U_{* \mathrm{max}}}{t_{\mathrm{min}}^3} \frac{L^6}{R_*^6}, 
\end{equation}
\noindent where $U_{* \mathrm{max}}$ corresponds to the maximum $U_*$ detectable for a given $A_{\mathrm{thresh}}$ of a star. Calculating this for the appropriate $\rho$ and $v_\mathrm{c}$, we arrive at
\begin{align}
\Gamma \approx & 2.63\times 10^{20} \left(\frac{L}{1 \mathrm{kpc}}\right)^6  \left(\frac{R_\sun}{R_*}\right)^6  \left(\frac{M}{M_\sun}\right)^4  \left(\frac{1\mathrm{hr}}{t_{\mathrm{min}}}\right)^3 \nonumber
\\ & \times \frac{1}{\left(A_{\mathrm{thresh}}^2-1\right)^5} \frac{1}{\mathrm{year}}.
\label{eq:approx}
\end{align}

\begin{figure}[htb!]
\begin{center}
\includegraphics[scale=.5,trim=0.5in 0 0 0]{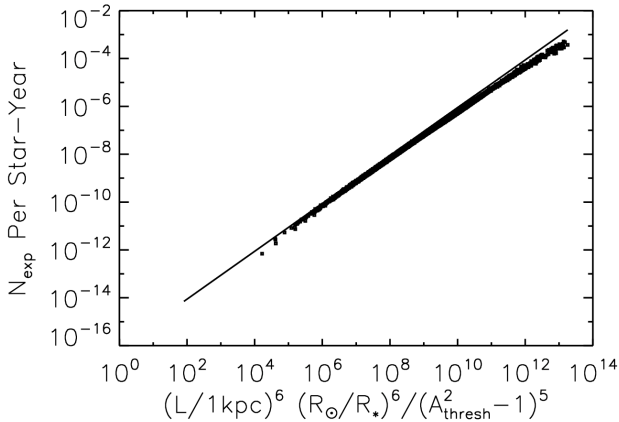}
\caption{Expected number of events per star-year for a PBH mass of $10^{-10} M_\sun$ with a 1 min cadence, for each non-variable dwarf star being monitored in the third quarter of the Kepler data, plotted with respect to the stellar variables governing this number. The straight line represents the approximation in Equation 25.}
\label{fig:Fig10}
\end{center} 
\end{figure}

\noindent Using the third quarter Kepler stars, we plot this approximation as a straight line in Figure \ref{fig:Fig10} along with the actual rates calculated using our integral formulas. The approximation works very well for these low-mass PBHs and demonstrates how the stellar characteristics come into play in this calculation. As seen in Equation \ref{eq:approx}, a high stellar distance-to-radius ratio is important. The dependence of the rate on the effective temperature is also readily explained by the fact that the distance is directly calculated from this value using $L=1.19\times10^{-3} R_* (T_{\mathrm{eff}}/T_\sun)^2 10^{0.2(V-A_\mathrm{V}+B.C.)}$ where $V$ is the apparent visual magnitude, $A_\mathrm{V}$ the extinction parameter, and B.C. the bolometric correction as in Paper I. We can see that the rate will be related to the effective temperature as $\Gamma \propto T_{\mathrm{eff}}^{12}$. Also, decreasing the cadence will have a cubic effect on the rate expected, while maintaining a low $A_{\mathrm{thresh}}$ is also important, as expected. We can therefore predict, for example, that for a Kepler type mission, with 30 min cadence, in order to push the PBH mass limits down to $5\times10^{-11} M_\sun$, one would have to observe 160,000 dwarf stars with a $(L/1 \mathrm{kpc})(R_\sun/R_*)$ fraction of 3.15 or higher, while maintaining the 0.1\% lightcurve precision. It seems therefore that pushing to smaller PBH masses in order to close the remaining PBH DM mass window will be difficult. However, if events were detected, a survey pointing towards or away from the Galactic center would provide us with more information about the DM distribution. We therefore address the WFIRST mission next.

\subsection{WFIRST}

\begin{figure}[htb!]
\begin{center}
\includegraphics[scale=.5,trim=0.5in 0 0 0]{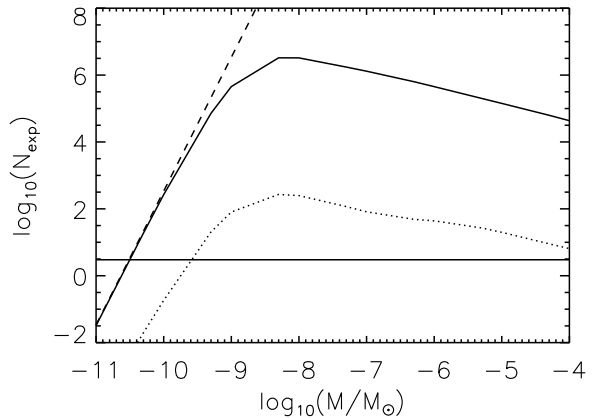}
\caption{Total number of expected events towards the galactic bulge for a mission such as WFIRST. The dashed line represents the approximation of Equation 25, appropriate for PBH lens masses below $10^{-10} M_\sun$. The horizontal line shows the 95\% confidence level limit if no events are detected. The dotted line represents the predicted Kepler microlensing limits as shown in Figure 1, for comparison.}
\label{fig:Fig11}
\end{center} 
\end{figure}

NASA's Wide-Field Infrared Survey Telescope (WFIRST), is a proposed next-generation space observatory being designed to search for Dark Energy and extrasolar planets \citep{Green2011}. It is designed to perform Dark Energy measurements using Baryon Acoustic Oscillations, Type Ia Supernovae, and Weak Lensing. In addition, it will complement the Kepler mission with its microlensing search for extrasolar planets, targeting stars towards the galactic bulge. Here we address its value for PBH DM limits or characterization. The preliminary specifications are to monitor $2 \times 10^8$ stars with a cadence of 15 min and a $1\%$ photometry precision \citep{Bennett2010}. It is especially exciting, as it will monitor the center of the Galaxy, and therefore could potentially provide insights into the DM distribution of the Milky Way. It will add to the existing DM dynamical constraints due to microlensing \citep{Iocco2011}.
Here we provide a preliminary calculation for the number of expected events, if the stars being monitored are similar to those of Kepler, as well as assuming a simple DM distribution of the form 

\begin{equation}
\rho(x)=\rho_0 \frac{a^2 +r^2_0}{a^2+L^2 (1-x)^2 },
\end{equation}
with $a = 5$ kpc. Following our analysis for the Kepler mission, we find the rate of detection,

\begin{equation}
\frac{d\Gamma}{dt_\mathrm{e}}=\rho_0 \frac{a^2 +r^2_0}{M} L v_c^2 \int_0^{x_{\mathrm{max}}} dx \frac{1}{a^2+L^2 (1-x)^2 } \beta^2 g(\beta)
\end{equation}

\noindent with all parameters as defined in Paper I.  In Figure \ref{fig:Fig11} we plot the number of expected events for one year of monitoring of Kepler-type stars (with $R_*\approx1 R_\sun$) toward the galactic bulge ($L=8$ kpc), with $25\%$ of stars assumed to be variable, and ignoring transverse velocity. Alongside, we plot the approximation given in Equation \ref{eq:approx} as a dashed line, demonstrating its usefulness for predicting microlensing rates for PBH lenses with masses less than $10^{-10} M_\sun$, as stated in the previous section.

This is a preliminary analysis, however, it demonstrates that WFIRST will complement Kepler in the same PBH mass range, providing a greater number of events, and exploring an additional order of magnitude of the PBH DM window. If the stars being monitored are larger than the Kepler type stars, the number of expected events will decrease, as predicted by Equation \ref{eq:approx}. Also, the transverse velocity of the source stars will have a sizable effect, as well as lensing due to other stars. If a PBHs are detected by Kepler, WFIRST will be able to study the DM characteristics, such as velocity and spatial distributions.

\section{Conclusions}

In this theoretical paper, we re-addressed the possibility of detecting or ruling out PBH DM using the microlensing of Kepler source stars. We introduced a more proper treatment of the population of source stars and their variability, including a finite-source microlensing framework which includes limb-darkening. Using this analysis, we found that the PBH DM mass constraints could be extended down to $2 \times 10^{-10} M_\sun$, two orders of magnitude below current limits and a third of a magnitude lower than our previous work. As mentioned, a proper Monte Carlo simulation will be needed to fully understand the experimental detection efficiency. We provide a limb-darkened microlensing framework which will be used to fit potential Kepler light curves, and which will help to distinguish between PBH's and stellar flares, the main source of experimental systematic error. Our analysis leaves us optimistic in being able to provide a probability for the masses of the lenses if any microlensing events are found, and therefore characterizing the DM and its epoch of formation. We found a very strong correlation between the rate of PBH detection for a given star and its effective temperature, providing an approximate expression for this rate for these low-mass PBHs. This should help in selecting stars to be monitored in future microlensing experiments. Using this approximation, it can be seen that fully closing the remaining PBH DM window using a microlensing method will be difficult, however sensitivity could potentially be improved by another order of magnitude using future planned missions, such as the WFIRST survey towards the galactic bulge. More analysis for this case is needed.

\acknowledgments

We thank Matthew J. Lehner and Alexander J. Mendez for helpful suggestions and discussions. K.G. and A.M.C. were supported in part by the DoE under grant DE-FG03-97ER40546. A.M.C. was supported in part by the National Science Foundation Graduate Research Fellowship under grant number DGE0707423. Some of the data presented in this paper were obtained from the Multimission Archive at the Space Telescope Science Institute (MAST). STScI s operated by the Association of Universities for Research in Astronomy, Inc., under NASA contract NAS5-26555. Support for MAST for non-HST data is provided by the NASA Office of Space Science via grant NNX09AF08G and by other grants and contracts.



\bibliographystyle{apj}
\bibliography{msbib}

\appendix

\end{document}